\documentstyle[aps,epsfig]{revtex}

\begin{document}
 
\title{Convergence to the critical attractor of dissipative maps: 
Log-periodic oscillations, fractality and nonextensivity} 
\vspace{2.5cm}

\author{F.A.B.F. de Moura}\address{Departamento de F\'{\i}sica,
Universidade Federal de Pernambuco, 50670-901 Recife - PE, Brazil}
\vspace{2cm}
\author{U. Tirnakli\thanks{e-mail: tirnakli@sci.ege.edu.tr}}
\address{Department of Physics, Faculty of Science,
Ege University 35100 Izmir, Turkey \\
and \\
Centro Brasileiro de Pesquisas F\'{\i}sicas, 
Rua Xavier Sigaud 150, 22290-180 Rio de Janeiro - RJ, Brazil}
\vspace{2cm}
\author{M. L. Lyra}\address{Departamento de F\'{\i}sica, Universidade
Federal de Alagoas, 57072-970 Macei\'o - AL, Brazil}

\maketitle
\vspace{1cm}

\begin{abstract}
For a family of logistic-like maps, we investigate the rate of 
convergence to the critical attractor when an ensemble of initial 
conditions is uniformly spread over the entire phase space. 
We found that the phase space volume occupied by the ensemble $W(t)$ 
depicts a power-law decay with log-periodic oscillations reflecting 
the multifractal character of the critical attractor. 
We explore the parametric dependence of the power-law exponent and the 
amplitude of the log-periodic oscillations with the attractor's 
fractal dimension governed by the inflexion of the map near its 
extremal point. Further, we investigate the temporal evolution of 
$W(t)$ for the circle map whose critical attractor is dense. 
In this case, we found $W(t)$ to exhibit a rich pattern with a slow 
logarithmic decay of the lower bounds. These results are discussed 
in the context of non-extensive Tsallis entropies. 

Pacs Numbers: 05.45.Ac, 05.20.-y, 05.70.Ce

\end{abstract}


\vspace{1cm}

\section{introduction}

Nonlinear low-dimensional dissipative maps can describe a great variety 
of systems with few degrees of freedom. The underlying nonlinearity can 
induce the system to exhibit a complex behavior with quite structured 
paths in the phase space. The sensitivity to initial conditions is a 
relevant aspect associated to the structure of the dynamical attractor.
 In general, the sensitivity is  measured as  the 
effect of any uncertainty on the system's variables. For systems 
exhibiting periodic or chaotic orbits, the effect of any uncertainty 
on initial conditions depicts an exponential temporal evolution with 
$\xi (t) \equiv\lim_{\Delta x(0)\rightarrow 0} \Delta x(t)/\Delta x(0) 
\sim e^{\lambda t}$, where $\lambda$ is the Lyapunov exponent, $\Delta
x(0)$ and $\Delta x(t)$ are the uncertainties at times $0$ 
and $t$. When the Lyapunov exponent $\lambda < 0$, $\xi (t)$ 
characterizes the rate of contraction towards periodic orbits. On the 
other hand, for $\lambda >0$, it characterizes the rate of divergence 
of chaotic orbits. 
At bifurcation and critical points  (i.e., onset to chaos) the 
Lyapunov exponent $\lambda$ vanishes. Recently, it was shown that this 
feature is related to a power-law sensitivity to initial conditions on 
the form\cite{Zheng,Lyra1,Lyra2}

\begin{equation}
\xi (t) = [1+(1-q)\lambda_qt]^{1/(1-q)}~~~,
\end{equation}
with $\lambda_q$ defining a characteristic time scale after which the
power-law behavior sets up.

A quantitative way to measure the sensitivity to initial conditions is 
to follow, from a particular partition of the phase space, the temporal 
evolution of the number of cells $W(t)$  occupied by an ensemble of
identical copies of the system. 
For periodic and chaotic orbits  $W(t) = W(0)e^{\lambda t}$.
In the particular case of equiprobability, the well-known Pesin equality 
reads $K=\lambda$ if $\lambda\ge 0$\cite{pesin} 
with $K$ being the
Kolmogorov-Sinai entropy \cite{KS}  defined as the 
variation per unit time of the standard Boltzmann-Gibbs entropy
$S=-\sum p_i\ln{p_i}$. 
This equality provides a link between the sensitivity to initial
conditions and the dynamic evolution of the relevant entropy. 

At bifurcation and critical points and for an ensemble of initial
conditions concentrated in a single cell, i.e. $W(0)=1$, it has been
shown that 
\begin{equation}
W(t) = [1+(1-q)K_qt]^{1/(1-q)}~~~,
\end{equation}
with $K_q$ being the generalized Kolmogorov-Sinai entropy  
defined as the rate of variation of the non-extensive Tsallis 
entropy $S_q=(1-\sum p_i^q)/(q-1)$\cite{Tsallis}.
The Pesin equality can be generalized
as  $K_q=\lambda_q$ if $\lambda_q \ge 0$\cite{Zheng}. 
Tsallis entropies have been successfully applied to recent studies of a 
series of non-extensive systems and provided a theoretical 
background to the understanding of some of their unusual physical 
properties\cite{bjp,html}.

The expansion towards the critical attractor
of an ensemble of initial conditions concentrated around the inflexion 
point of the map can be characterized by a proper  $S_q$ evolving at a
constant rate. Scaling arguments have shown that the appropriate 
entropic index $q$ is related to the multifractal
structure of the critical dynamical attractor by\cite{Lyra2}
\begin{equation}
\frac{1}{1-q} = \frac{1}{\alpha_{min}}-\frac{1}{\alpha_{max}}
\end{equation}
where $\alpha_{min}$ and $\alpha_{max}$ are the extremal singularity
strengths of the multifractal spectrum of the critical
attractor\cite{Halsey}. The above scaling relation has been shown to 
hold for the families of generalized logistic and circle
maps\cite{Lyra2,Lyra3,Lyra4,Lyra5}.

However, the temporal evolution of critical dynamical systems can be 
strongly dependent on the particular initial ensemble. Although some
scaling laws can be found for an ensemble of initial conditions
concentrated around the map inflexion point, these are usually not
universal with respect to a general ensemble. In this work, we are going
to numerically investigate the critical temporal evolution of the volume
of the phase space occupied by an ensemble of initial conditions spread
over the entire phase space. This ensemble is expected to contract towards
the critical attractor. Using a family of 
one-dimensional generalized logistic maps having $d_f<1$, we will perform
a detailed study of the  parametric dependence of 
$W(t)$ on the fractal dimension of the critical attractor. 
Due to the discrete scale invariance of the critical
attractor, the convergence displays log-periodic oscillations\cite{Sornette}.
We are also going to explore the dependence of the amplitude of these
oscillations with respect to the attractor's fractal dimension.  Further,
the behavior of $W(t)$ will be investigated  for the one-dimensional
critical circle map having $d_f=1$. For this map, the temporal evolution
is expected to display distinct trends since the critical 
attractor is dense.

\section{The convergence to the critical attractor of generalized
logistic maps}

Logistic-like maps are the simplest one-dimensional nonlinear dynamical
systems which allow a close investigation of a series of critical
exponents related to the onset of chaotic orbits. This family reads
\begin{equation}
x_{t+1}=1-a|x_t|^z~~~;~~~(z>1~~;~~0<a<2~~; t=0,1,2,...~~; 
x_t\in [-1,1])~~.
\end{equation}
Here $z$ is the inflexion of the map in the neighborhood of the extremal
point $\bar{x}=0$. These maps are well known to have topological
properties not dependent of $z$. However, the metrical properties, such as
Feigenbaum exponents\cite{Feigenbaum,Coullet} and the multifractal
spectrum of the critical attractor do depend on $z$. In particular the
fractal dimension of the critical attractor $d_f(z) <1$\cite{Beck} and
therefore it does not fill a finite fraction of the phase space. For a set
of initial conditions spread in the vicinity of the inflexion point, it
was found that the volume in phase space occupied by the ensemble grows
following a rich pattern with the upper bounds $W_{max}(t)$ governed by a
power-law $W_{max}(t)\propto t^{1/(1-q)}$, where $q$ is the entropic index
characterizing the relevant Tsallis entropy that grows at a constant 
rate. It has been shown that the dynamic exponent $1/(1-q)$ is directly
related to geometric scaling exponents related to the extremal sets of
the dynamic attractor\cite{Lyra2}.

Due to the presence of long-range spatial and temporal correlations at
criticality, one expects  the critical exponent governing the temporal
evolution to be sensitive to the particular initial ensemble. Indeed,
the multifractal spectrum characterizing the critical dynamical attractor
indicates that an infinite set of exponents are needed to fully
characterize the scaling behavior. In particular, an ensemble consisting
of a set of identical systems whose initial conditions is spread over the
entire phase-space is a common one when studying non-linear as well as  
thermodynamical systems.

Here, we will follow the dynamic evolution, in phase space, of an
ensemble of initial conditions uniformly distributed over the phase-space
and explore its relation with the generalized fractal dimensions of the
critical attractor.  In practice, a partition of the phase space on
$N_{box}$ cells of equal size is performed and a set of $N_c$ identical
copies of the system is followed whose initial conditions are uniformly
spread over the phase-space. The ratio $r=N_c/N_{box}$ is a control 
parameter giving the degree of sampling of the phase-space.

Within the non-extensive Tsallis statistics, there is a proper entropy $S_q$
 evolving at a constant rate such that
\begin{equation}
K_q = \lim_{N_{box}\rightarrow\infty} [S_q(t) - S_q(0)]/t
\end{equation}
goes to a constant value as $t\rightarrow\infty$. Notice that $K_q<0$ for
the process of convergence towards the critical attractor. Assuming that
all cells of the partition are occupied with equal probability, the
entropy $S_q(t)$ can be written as
\begin{equation}
S_q(t) = \frac{1-\sum_{i=1}^{W(t)} p_i^q}{q-1} = \frac{W(t)^{1-q}-1}{1-q}
\end{equation}
The last two equations imply that the number of occupied cells evolves in
time as
\begin{equation}
W(t) = [W(0)^{1-q}+(1-q)K_qt]^{1/(1-q)}
\end{equation}
with the exponent $\mu = -1/(q-1) >0 $ governing the asymptotic
power-law decay.

In figure~1, we show our results for $W(t)/N_{box}$ in the standard
logistic map with inflexion $z=2$ and from distinct
partitions of the phase space with sampling ratio $r=0.1$ . We observe
that, after a short transient period when $W(t)$ is nearly constant, a
power-law contraction of the volume occupied by the ensemble sets up.
$W(t)$ saturates at a finite fraction corresponding to the phase space
volume occupied by the critical attractor on a given finite partition. The
saturation is postponed when a finer partition is used once the fraction
occupied by the critical attractor vanishes in the limit
$N_{box}\rightarrow\infty$.

In figure~2, we show $W(t)/N_{box}$ for a given fine partition of the
phase-space and distinct sampling ratios $r$. We notice that the crossover
regime to the power-law scaling is quite short for large values of $r$ so
that a clear power-law scaling regime sets up even at early times. This 
feature is consistent with eq.(7) which states that the crossover time
$\tau$ scales as $\tau\sim 1/W(0)^{q-1}$. 
Further, the scaling regime exhibits log-periodic oscillations once the
multifractal nature of the critical attractor is closely probed by such
dense ensemble. A general form for $W(t)$ reflecting the discrete scale  
invariance of the attractor can be written as

\begin{equation}
W(t) = t^{-\mu}P\left( \frac{\ln{t}}{\ln{\lambda}}\right)
\end{equation}
where $P$ is a function of period unity and $\lambda$ is the 
characteristic scaling factor between the periods of two consecutive
oscillations. These log-periodic
oscillations have been observed in a large number of systems exhibiting
discrete scale invariance\cite{Sornette}. In general the amplitude of
these oscillations ranges form $10^{-4}$ up to $10^{-1}$.
Keeping only the first term in a Fourier series of
$P(\ln{t}/\ln{\lambda})$, one can write $W(t)$ in the form
\begin{equation}
W(t) = c_0t^{-\mu}\left[ 1 +
2\frac{c_1}{c_0}
\cos{\left(2\pi\frac{\ln{t}}{\ln{\lambda}} +\phi\right)}\right]\;\; .
\end{equation}

Log-periodic modulations correcting a pure power-law have been found
in several systems, as, for example, 
diffusion-limited-aggregation\cite{Sornette2}, crack growth\cite{Ball},
earthquakes\cite{Newman} and financial markets\cite{Drozdz}. 
It has also been observed in thermodynamic systems with fractal-like
energy spectrum\cite{Mendes,Vallejos}. The factors
controlling the log-periodic relative amplitude $2c_1/c_0$ are not well
known for most of the systems where it has been observed. In the present
study, we can closely investigate the factors which may control these
amplitudes by measuring it as a function of the map inflexion $z$ for a
fixed partition and sampling ratio (see figure 3). We found that these
oscillations have amplitudes decaying exponentially with $z$  as shown in
figure~4. It is interesting to point out that the fractal dimension of
the attractor is a monotonically decreasing function of $z$. Therefore,
the above trend indicates a possible correlation between the
amplitude of the log-periodic oscillations and the fractal dimension of
the dynamical attractor.

We also measured the critical exponent $\mu$ as a function of the map
inflexion $z$. Our results are summarized in the Table. It is a decreasing
function of $z$ as can be seen in figure~5. The volume occupied by the
ensemble depicts a fast contraction for $z\sim 1$ where the fractal
dimension is small.  On the other side, a very slow contraction is
observed for large values of $z$, pointing towards a saturation or at most
to a logarithmic decrease of $W(t)$ in the limit of dense attractors. We
would like to point out here that the exponent governing the expansion
of the volume occupied by an ensemble of initial conditions concentrated
around the inflexion point exhibits a reversed trend. Although scaling
arguments have shown that this exponent can be written in terms of
scaling exponents characterizing the extremal sets in the attractor, we
could not devise a simple scaling relation between $\mu$ and the
multifractal singularity spectrum. However, we observed
that, when plotted against the fractal dimension of the attractor as
shown in figure~6, the dynamic exponent $\mu$ is very well fitted by
$\mu\propto (1-d_f)^2$, which indicates $d_f$ as the relevant geometric
exponent coupled to the dynamics of the uniform ensemble. 
We would like to mention here that the 
same dynamic exponents were obtained for the generalized periodic maps
which belong to the same universality class of  logistic-like
maps\cite{Lyra3}.

\section{The convergence to the critical attractor of the circle map}

The results from the previous section indicate that a slow convergence to
the critical attractor shall be expected for dense critical attractors.
However, it is not clear in what fashion this convergence will take place
when the dynamical attractor fills the  phase space  with a
multifractal  probability density as occurs for the one-dimensional
critical circle map

\begin{equation}
\theta_{t+1} = \theta_t +\Omega -\frac{1}{2\pi}\sin{(2\pi\theta_t)}
~~~ mod(1)
\end{equation}
where $0\leq\theta_t<1$ is a point on a circle. The circle map describes
dynamical systems possessing a natural frequency $\omega_1$ which are
driven by an external force of frequency $\omega_2$ ($\Omega =
\omega_1/\omega_2$ is the bare winding number) and belongs to the same
universality class of the forced Rayleigh-B\'enard 
convection\cite{Jensen}. For $\Omega = 0.606661...$ the circle map has a
cubic inflexion ($z=3$) in the vicinity of the point $\bar{\theta}=0$.
Starting from a given point on the circle, it generates a quasi-periodic
orbit which fills the  phase-space and the dynamical attractor is
a multifractal with fractal dimension $d_f=1$\cite{Halsey}.

In figure~7 we show our results for the temporal evolution of the
phase-space volume occupied by an ensemble of initial conditions uniformly
spread over the circle. $W(t)$ exhibits a rich pattern which resembles the
one observed for the sensitivity function associated to the expansion of
the phase-space from initial conditions concentrated around the inflexion
point. However, $W(t)$ does not present any power-law regime. Instead, the
lower bounds display a slow logarithmic decrease with time, saturating at
a finite volume fraction. The saturation is a feature related to the
finite partition used in the numerical calculation. This minimum
decreases logarithmically with the number of cells in the phase-space as 
shown in figure~8. We also observed the same behavior for generalized
circle maps with an arbitrary inflexion $z$\cite{Lyra5}. The critical
attractors within this family have all $d_f=1$ although 
they exhibit a $z$-dependent multifractal singularity spectra. 
The $z$-independent scenario for $W(t)$ corroborates the 
conjecture 
that $d_f$ is the relevant geometric exponent coupled to the dynamics 
of the uniform ensemble.

\section{Summary and Conclusions}

In this work, we studied the temporal evolution in phase space of an
ensemble of identical copies of one-dimensional nonlinear dissipative
maps. We found that the phase-space volume occupied by an initially
uniform ensemble displays a power-law decay with log-periodic oscillations
whenever the dynamical attractor has a fractal dimension $d_f<1$. The
amplitude of the oscillations was found to depict a monotonic parametric
dependence on $d_f$. For dense multifractal attractors, $W(t)$ presents 
only a slow logarithmic contraction of its lower bounds followed by a
rich pattern.

The critical exponent characterizing the contraction of the uniform
ensemble was found to have no direct relation to the one governing the
expansion from a set of initial conditions concentrated around the
inflexion point. In particular, no power-law was found for the contraction
in the standard and generalized circle maps, in contrast to the
$z$-dependent power-law expansion. These results indicate that the
relevant Tsallis entropy evolving at a constant rate (modulated by
log-periodic oscillations) is characterized by an entropic index $q$
that depends on the initial ensemble.  It would be valuable to
investigate the possible existence of classes of ensembles with a common
dynamics in phase-space and, therefore, characterized by the same
entropic index $q$. The non-universality of $q$ with respect to the
initial ensemble is related to the multifractal character of the dynamical
attractor. However, as for the ensemble concentrated at the vicinity of
the inflexion point, the exponent governing the dynamics of the uniform
ensemble is  coupled to a geometric scaling exponent, in particular to the
proper fractal dimension of the attractor. This result comes in favor of
the concept that the degree of nonextensivity of the entropy measure
evolving at a constant rate is related to the fractal nature of the
dynamical attractor.

\section{acknowledgments}
UT acknowledges the partial support of BAYG-C program of TUBITAK 
(Turkish agency) as well as CNPq and PRONEX (Brazilian agencies).
This work was partially supported by CNPq and CAPES (Brazilian research
agencies). 
MLL would like to thank the hospitality of the Physics Department at
Universidade Federal de Pernambuco during the Summer School 2000 where
this work was partially developed.

\newpage


\newpage

\section*{FIGURE AND TABLE CAPTIONS}

\noindent
{\bf Figure 1 -} The volume occupied by the ensemble $W(t)$ as a function
of time in the standard logistic map ($z=2$) and with sampling ratio
$r=0.1$. From top to bottom $N_{box} = 2000, 8000, 32000, 128000$.

~

\noindent
{\bf Figure 2 -} The volume occupied by the ensemble $W(t)$ as a function
of time in the standard logistic map ($z=2$) and for a partition
containing $N_{box} = 128000$ cells. Notice the emergence of log-periodic
oscillations for large sampling ratios.

~

\noindent
{\bf Figure 3 -} The periodic function $W(t)/(c_0t^{-\mu})$ versus time
within the scaling regime and for $r=10$. Data from map inflexions 
$z=1.1, 1.25, 1.5, 2.0$ are shown. The amplitude of the oscillations
decreases monotonically as $z$ increases, but the characteristic scaling
factor between the periods of two consecutive  oscillations
is roughly $z$-independent.

~

\noindent
{\bf Figure 4 -} The amplitude of the log-periodic oscillations 
$2c_1/c_0$ as a function of the map inflexion $z$ for sampling ratio 
$r=10$. The monotonic decrease of the oscillations indicates a close 
relation between these and the fractal dimension of the underlying 
dynamical attractor.

~

\noindent
{\bf Figure 5 -} The dynamic exponent $\mu$ governing the contraction of
the occupied phase space volume [$W(t)\propto t^{-\mu}$] as a function of  
the map inflexion $z$. 

~

\noindent
{\bf Figure 6} - The parametric dependence of the dynamic exponent $\mu$
with the fractal dimension $d_f$ of the critical attractor. It is very
well fitted to the form $\mu\propto (1-d_f)^2$, indicating that $d_f$
seems to be the relevant geometric exponent coupled to the dynamics of the
uniform ensemble.

~

\noindent
{\bf Figure 7 -} The volume occupied by the ensemble $W(t)$ as a 
function of time in the standard critical circle map. The lower bounds 
display a slow logarithmic decay with time saturating at a finite volume
fraction due to the finite partition of the phase space.

~

\noindent
{\bf Figure 8 -} The asymptotic lower bounds for the occupied volume in
the phase space versus the number of cells $N_{box}$. The logarithmic
decay agrees with the  prediction that $\mu (d_f\rightarrow 1)\rightarrow
0$. The same behavior was observed for the family of generalized circle 
maps and corroborates the conjecture that $d_f$ is the relevant
geometric exponent coupled to the dynamics of the uniform ensemble.

~

\noindent
{\bf Table } -  Numerical values, within the $z$-generalized family of 
logistic maps, of: {\em i}) the dynamic exponent $\mu$ governing the 
contraction towards the critical attractor of the uniform ensemble; 
{\em ii}) the entropic index $q$ of the proper Tsallis entropy 
decreasing at a constant rate; {\em iii}) the fractal dimension $d_f$ 
of the critical attractor. These values also hold for the generalized 
periodic maps. The last line represents our results for the 
$z$-generalized circle maps.

\newpage
\begin{center}
{\bf Table}

\vspace{1.52cm}

\begin{tabular}{||c|c|c|c||} \hline
$z$ & $~\mu = -1/(1-q)~$ & $q$ &  $d_f$ \\ \hline
$1.10$ & $1.62 \pm 0.02$  & $1.62\pm 0.01$ & $~0.32 \pm 0.02~$ \\ \hline
$1.25$ & $1.23 \pm 0.01 $ &  $1.81\pm 0.01$ & $0.40\pm 0.01$ \\ \hline
$1.5$ &  $0.95 \pm 0.01$ &  $2.05\pm 0.01$ & $0.47\pm 0.01$ \\ \hline
$1.75$ & $0.80 \pm 0.01$ &  $~2.25\pm 0.015~$ & $0.51\pm 0.01$ \\ \hline
$2.0$ & $ 0.71 \pm 0.01$ &  $2.41\pm 0.02$ & $0.54\pm 0.01$ \\ \hline
$2.5$ & $0.59 \pm 0.01$ &  $2.70\pm 0.02$ & $ 0.58\pm 0.01$ \\ \hline
$3.0$ & $0.515 \pm 0.005$ & $2.94\pm 0.02$ &$ 0.60\pm 0.01$ \\ \hline
$5.0$ & $0.395 \pm 0.005$ & $3.53\pm 0.03$ &$ 0.66\pm 0.01$ \\ \hline
$z$-circular  &   &    & \\
maps & $ 0.0 $ & $\infty$ &$ 1.0 $ \\ \hline 
\end{tabular}
\end{center}

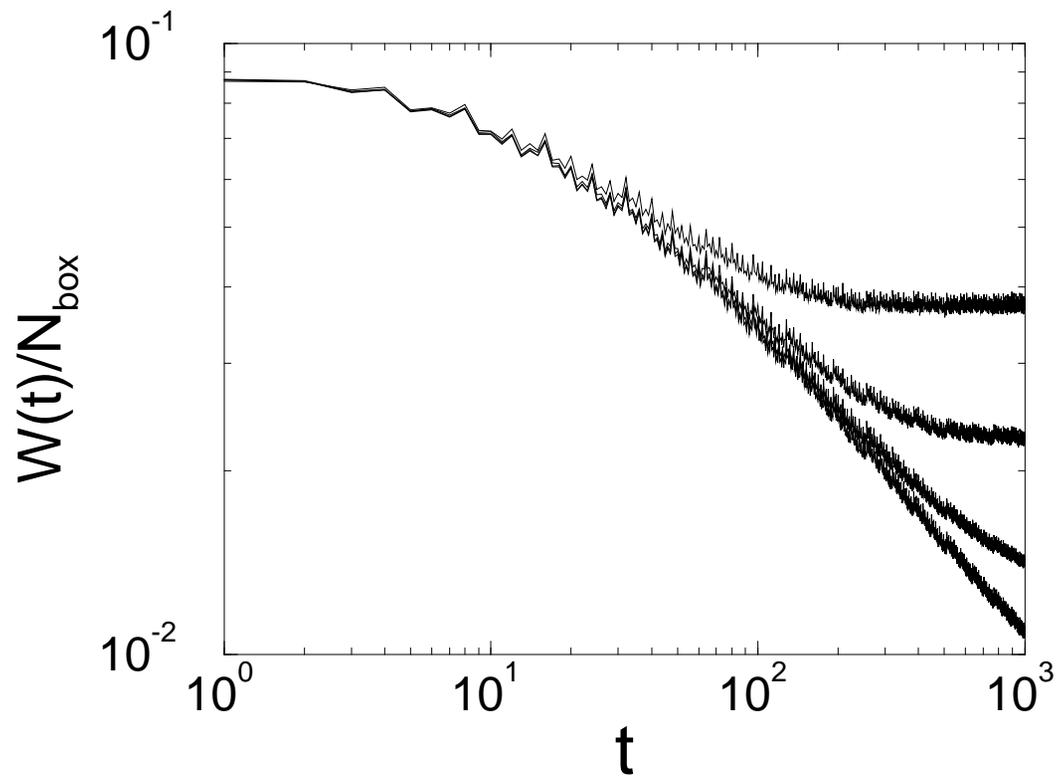
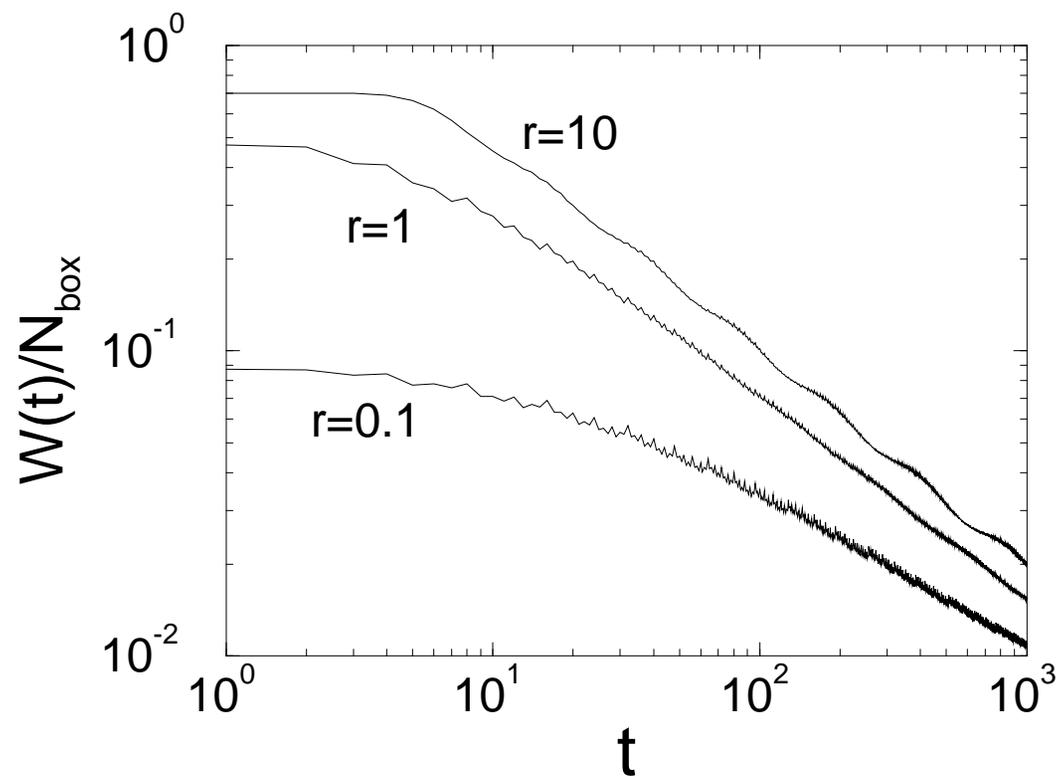
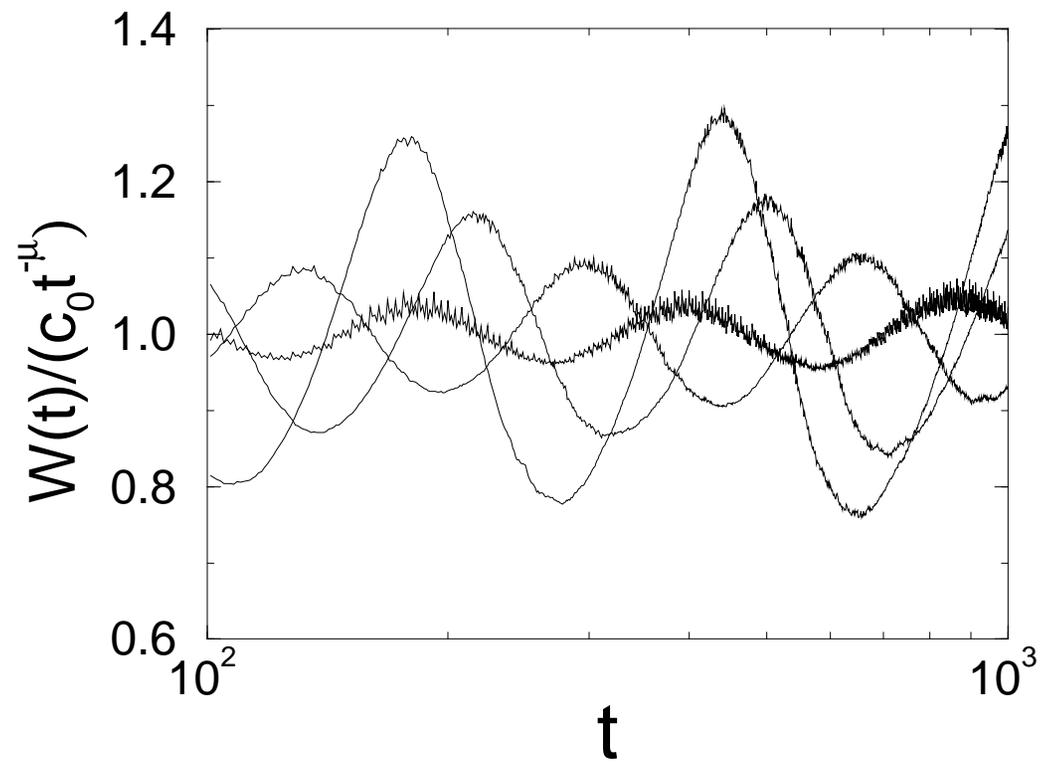
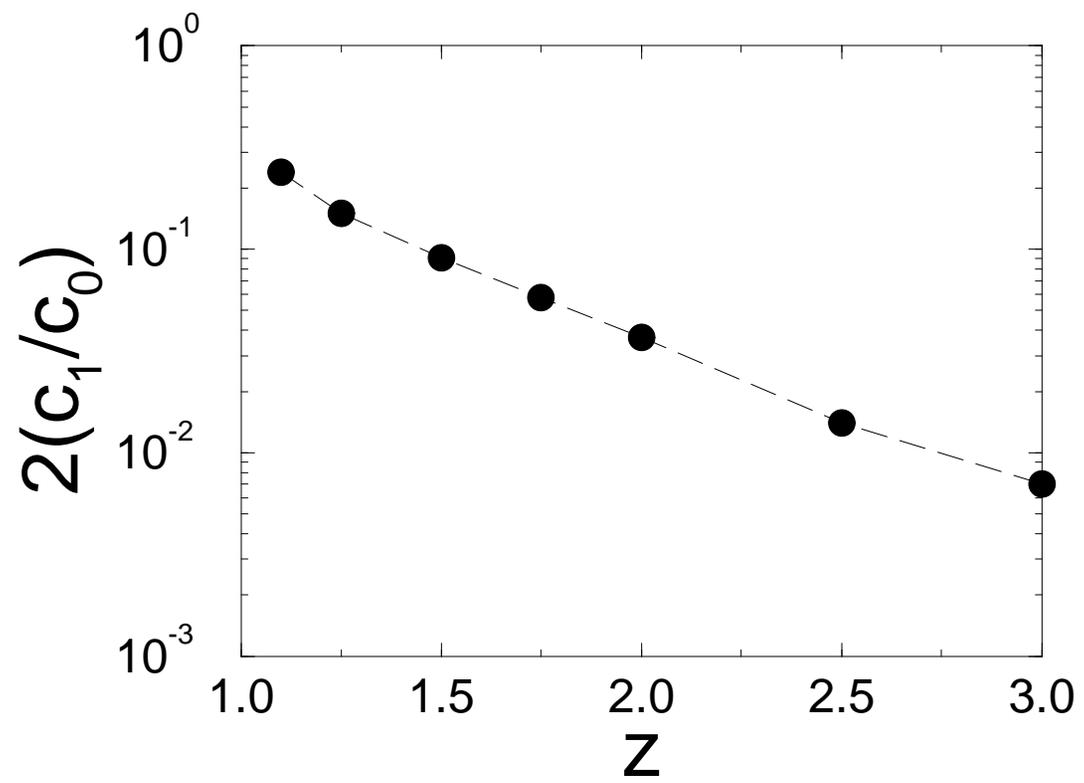
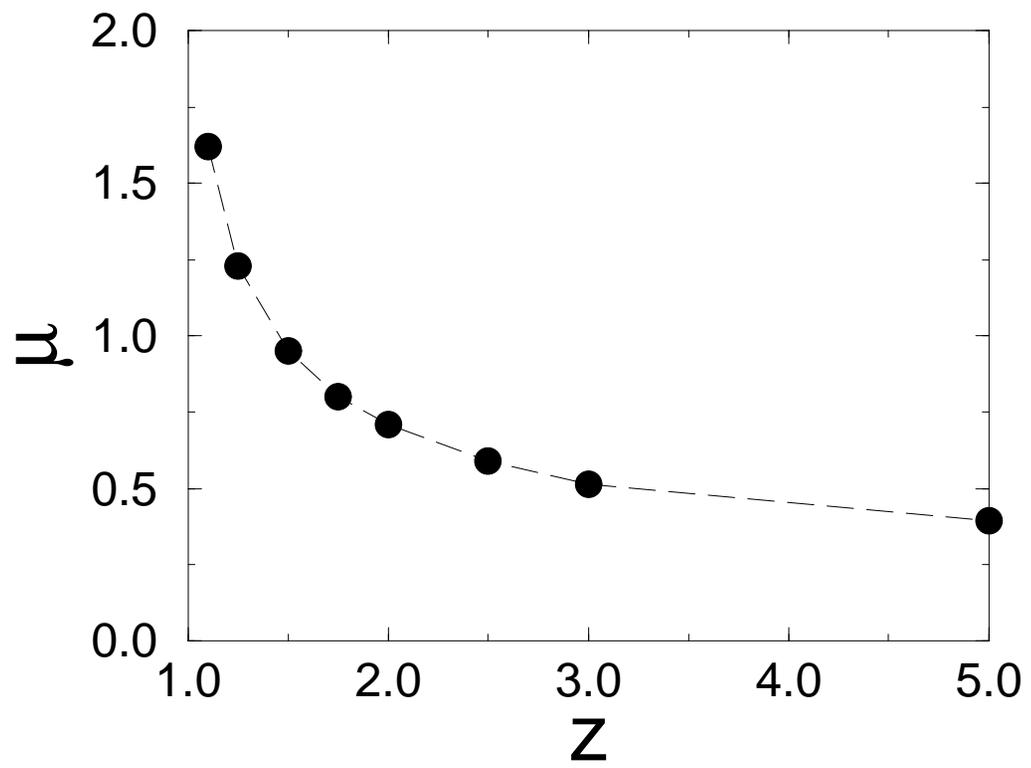
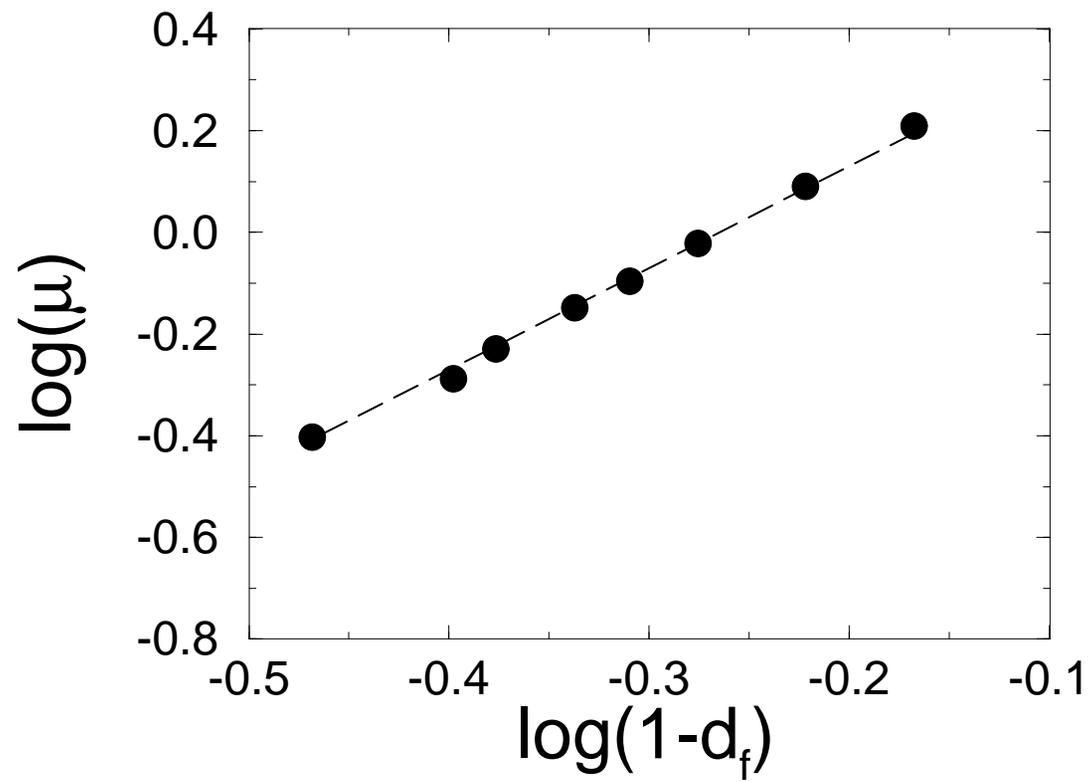
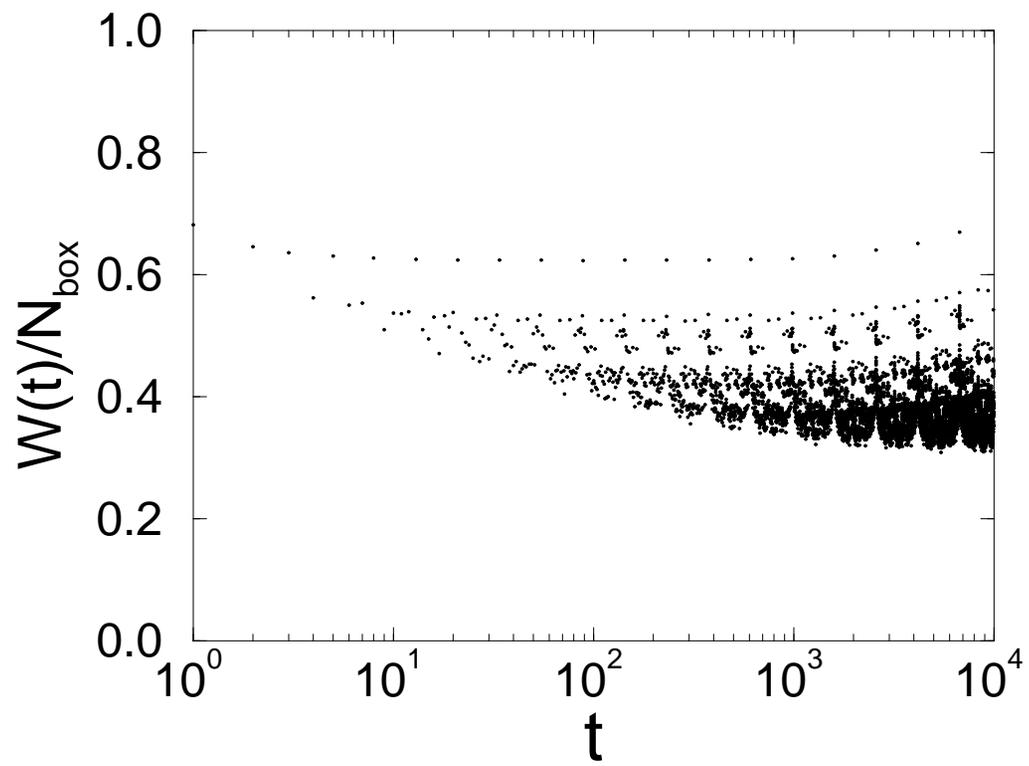
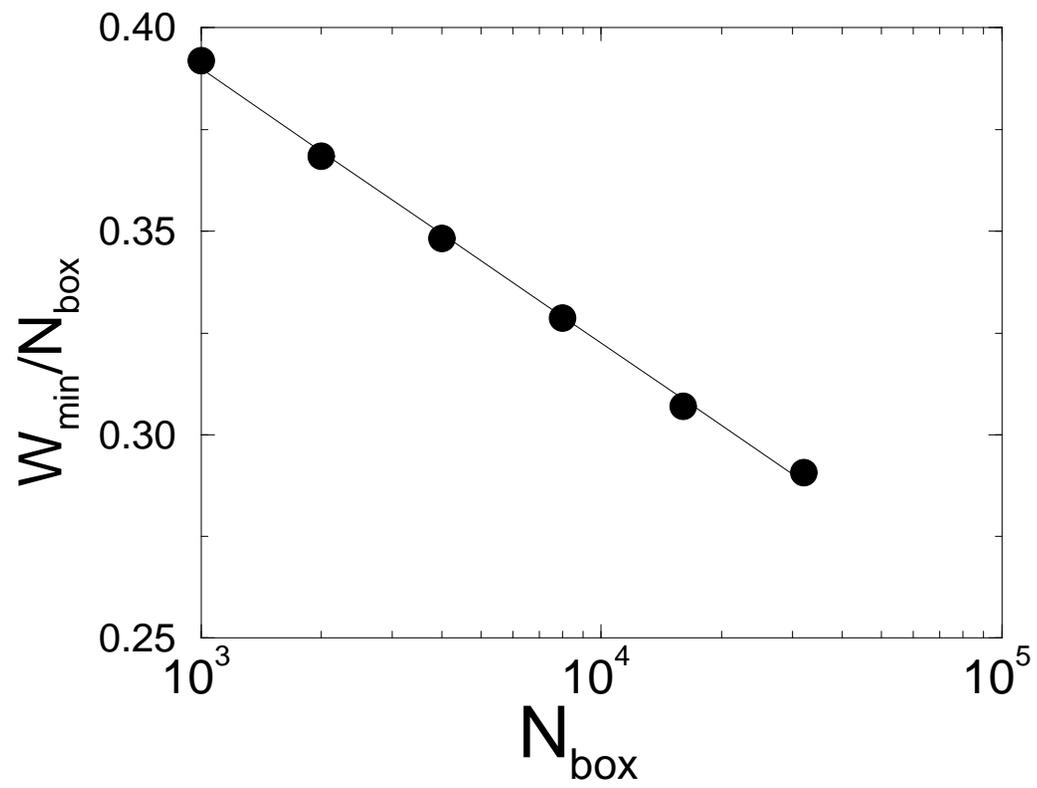
\end{document}